\documentclass[10pt,pre,showpacs]{revtex4}
\usepackage{graphicx}

\def\be{\begin{equation}}
\def\ee{\end{equation}}

\begin{document}
\title{STOCHASTIC MODEL OF THE CONDITIONAL LAGRANGIAN ACCELERATION
OF A FLUID PARTICLE IN DEVELOPED TURBULENCE}
 \author{A.K. Aringazin}
 \email{aringazin@mail.kz}
 \author{M.I. Mazhitov}
 \email{mazhitov@emu.kz}
 \affiliation{Department of Theoretical Physics, Institute for
Basic Research, Eurasian National University, Astana 473021,
Kazakhstan}

\date{November 5, 2003}

\begin{abstract}
{The random intensity of noise approach to 1D
Laval-Dubrulle-Nazarenko model is used to describe Lagrangian
acceleration of a fluid particle in developed turbulence.
Intensities of noises entering nonlinear Langevin equation are
assumed to depend on random velocity fluctuations in an
exponential way. The conditional acceleration PDF, variance, and
mean are found to be in a good qualitative agreement with the
recent high-Re Lagrangian experimental data.}
\end{abstract}


\maketitle



\noindent Phenomenological approaches~\cite{Aringazin,Beck3} were
used~\cite{Beck4-Beck} to describe Lagrangian acceleration of a
fluid particle in developed turbulent flow within the framework of
Langevin type equation; see
also~\cite{Sawford-Wilk-Beck2-Reynolds}. Recent one-dimensional
(1D) stochastic particle models and some
refinements~\cite{Aringazin2-Aringazin3,Aringazin4} were reviewed
in~\cite{Aringazin5}. Some toy models of developed turbulence
suffer from the lack of physical interpretation deduced from
turbulence dynamics~\cite{Kraichnan0305040}.

Recently~\cite{Aringazin5,Aringazin6} we have shown that the 1D
Laval-Dubrulle-Nazarenko (LDN) type toy model~\cite{Laval,Laval2}
of the acceleration evolution with the model turbulent viscosity
$\nu_{\mathrm t}$ and coupled delta-correlated Gaussian
multiplicative and additive noises is in a good agreement with the
recent high-precision Lagrangian experimental data on acceleration
statistics; $R_\lambda=690$, the normalized acceleration range is
$[-60,60]$, Kolmogorov scale is
resolved~\cite{Bodenschatz,Bodenschatz2,Mordant0303003}. The
longstanding Heisenberg-Yaglom scaling, $\langle a^2\rangle =
a_0{\bar u}^{9/2}\nu^{-1/2}L^{-3/2}$, was confirmed
experimentally~\cite{Bodenschatz} to a very high accuracy, for
about seven orders of magnitude in the acceleration variance, or
two orders of the {\em rms} velocity $\bar u$, at $R_\lambda>500$.
Long-time correlations and the occurrence of very large
fluctuations at small scales dominate the motion of a fluid
particle, and this leads to a new dynamical picture of
turbulence~\cite{Mordant,Chevillard0310105}. We focus on modeling
the acceleration statistics conditional on velocity fluctuations
$u$ presented recently in~\cite{Mordant0303003}.

The original 3D and 1D LDN models were formulated both in the
Lagrangian and Eulerian frameworks for small-scale velocity
increments in time and space respectively. They are based on a
stochastic kind of Batchelor-Proudman rapid distortion theory
approach to the 3D Navier-Stokes equation~\cite{Laval}, and thus
have a deductive support from turbulence dynamics. The random
intensity of noise (RIN) approach~\cite{Aringazin5,Aringazin6}
provides an extension of the above 1D model in the limit of small
time scale $\tau$ for which Lagrangian velocity increments are
proportional to $\tau$: $u(t+\tau)-u(t) = \tau a(t)$. The main
idea of RIN approach is simply to account for the recently
established two well separated Lagrangian autocorrelation time
scales for the velocity increments~\cite{Mordant} and assume that
certain model parameters, such as intensities of the noises,
fluctuate at the {\em long} time scale due to Lagrangian velocity
$u$. A simple 1D model can shed some light to properties of 3D LDN
model of Lagrangian dynamics.


We give only a brief sketch of the 1D LDN model and refer the
reader to~\cite{Laval,Aringazin5} for more details; see
also~\cite{Dubrulle0304035}. This toy model can also be viewed as
a passive scalar in a compressible 1D flow~\cite{Laval}. We use
probability density function (PDF) obtained as a stationary
solution of the Fokker-Planck equation associated to the Langevin
equation for the component of Lagrangian acceleration $a(t)$:
$\partial a/\partial t = (\xi - \nu_{\mathrm t}k^2)a +
\sigma_\perp$~\cite{Laval}. In 3D LDN model, $\xi(t)$ is related
to the velocity derivative tensor and $\sigma_\perp(t)$ describes
a forcing of small scales by large scales via the energy cascade
mechanism. In 1D LDN model, these are approximated by external
Gaussian white noises,
\be\label{noises}
\langle\xi(t)\rangle=0, \ \langle\xi(t)\xi(t')\rangle =
2D\delta(t-t'), \ \langle\sigma_\perp(t)\rangle = 0, \
\langle\sigma_\perp(t)\sigma_\perp(t')\rangle =
2\alpha\delta(t-t'), \ \langle\xi(t)\sigma_\perp(t')\rangle =
2\lambda\delta(t-t'),
\ee
that is partially justified by DNS~\cite{Laval}. The acceleration
PDF can be calculated exactly~\cite{Aringazin5}, with the result
\be\label{PLaval} P(a) = {C \exp[-{\nu_{\mathrm
t}k^2}/{D}+F(c)+F(-c)]} {(Da^2\!-\!2\lambda a
\!+\!\alpha)^{1/2}(2Bka+\nu_{\mathrm t}k^2)^{{-2B\lambda
k}/{D^2}}},
\ee
for constant parameters. Here, we have denoted $\nu_{\mathrm t} =
\sqrt{\nu_0^2+ B^2a^2/k^2}$, $C$ is normalization constant,
\begin{eqnarray}\label{A4}
F(c)
 = \frac{c_1k^2}{2c_2D^2c}\ln[\frac{2D^3}{c_1c_2(c-Da+\lambda)}
   (
   B^2(\lambda^2 + c\lambda-D\alpha)a
   + c(D\nu_{\mathrm t}^2k^2+c_2\nu_{\mathrm t})
    )
   ], \ c=-i\sqrt{D\alpha-\lambda^2}, \\
c_1 = B^2(4\lambda^3\!+\!4c\lambda^2\! -\!
3D\alpha\lambda-cD\alpha)
    \!+\! D^2(c\!+\!\lambda)\nu_0^2k^2,\
c_2 = \sqrt{B^2(2\lambda^2 + 2c\lambda-D\alpha)k^2 +
D^2\nu_0^2k^4}.
\end{eqnarray}


Without loss of generality one can put, in a numerical study,
$k=1$ and the additive noise intensity $\alpha=1$ by rescaling of
the multiplicative noise intensity $D>0$, the turbulent viscosity
parameter $B>0$, the kinematic viscosity $\nu_0>0$, and the cross
correlation $\lambda$~\cite{Aringazin5}, and make a fit of $P(a)$
to the experimental data. The particular cases $B=0$ and $\nu_0=0$
at $\lambda=0$ were studied in detail in~\cite{Aringazin5}.
Nonzero $\lambda$ is responsible for an asymmetry of the PDF
(\ref{PLaval}) and in 3D picture corresponds to a correlation
between stretching and vorticity (the energy cascade); in the
Eulerian framework, $\langle (\delta u)^3\rangle$ is proportional
to $\lambda$, in accord to a kind of Karman-Howarth
equation~\cite{Laval}.


Since the experimental unconditional and conditional
distributions, $P_{\mathrm{exp}}(a)$ and $P_{\mathrm{exp}}(a|u)$
at $u=0$, were found to be approximately of the same stretched
exponential form revealing strong Lagrangian
intermittency~\cite{Mordant0303003} we use the result of our
fit~\cite{Aringazin6} of the PDF (\ref{PLaval}) to
$P_{\mathrm{exp}}(a)$~\cite{Bodenschatz2} measured with 3\%
relative uncertainty for $|a|\leq 10$. This implies the following
rescaled parameter set: $D=1.100$, $B=0.155$, $\nu_0=2.910$,
$\lambda=-0.005$ ($k=1$, $\alpha=1$, $C=3.230$). A fit to the {\em
conditional} distribution $P_{\mathrm{exp}}(a|0)$ would yield a
different set of values of the parameters. The used fit is however
justified on a qualitative level. We assume that the parameters
$\alpha$ and $\lambda$ entering (\ref{PLaval}) depend on the
amplitude of (normalized) Lagrangian velocity fluctuations $u$,
while $D$, $B$, and $\nu_0$ are taken to be fixed at the fitted
values ($k=1$). An exponential form of $\alpha(u)$ has been
proposed in~\cite{Aringazin5} and was found to be relevant from
both the (K62) phenomenological and experimental points of view.
Particularly, such a form leads to log-normal RIN model when $u$
is independent Gaussian distributed with zero mean (PDF $g(u)$),
and yields the acceleration PDF whose low-probability tails are in
agreement with experiments~\cite{Beck4-Beck,Aringazin5}; the
marginal PDF is
$P_{\mathrm{m}}(a)=\int_{-\infty}^{\infty}P(a|u)g(u)du$.
Remarkably, this form is found to provide appreciable increase of
the {\em conditional acceleration variance} $\langle a^2|u\rangle$
with increasing $|u|$~\cite{Aringazin5} that meets the
experimental data~\cite{Mordant0303003}. Guided by these
observations the simplest choice is to try an exponential
dependence for $\lambda(u)$. Particularly, in the present
description we use $\lambda(u)=-0.005e^{3|u|}$ and
$\alpha(u)=e^{|u|}$. This allows us to deal with the acceleration
statistics which exhibits a strongly non-Gaussian character in
both the unconditional and conditional cases.
\begin{figure}[tbp!]
  \begin{minipage}[t]{.45\linewidth}
  \begin{center}
    \mbox{
    \includegraphics[height=6cm]{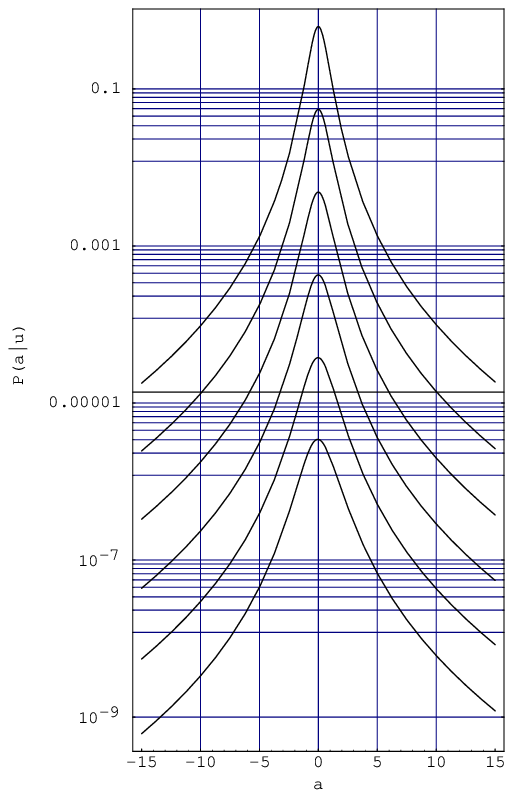}}
    \caption{\label{Fig1} The conditional acceleration PDF
$P(a|\alpha(u), \lambda(u))$ given by Eq.~(\ref{PLaval}). $u=$ 0
(top curve), 0.25, 0.50, 0.75, 1.00, 1.19 (bottom curve).}
  \end{center}
  \end{minipage}\hfill
  \begin{minipage}[t]{.45\linewidth}
  \begin{center}
    \mbox{
    \includegraphics[height=6cm,width=7.5cm]{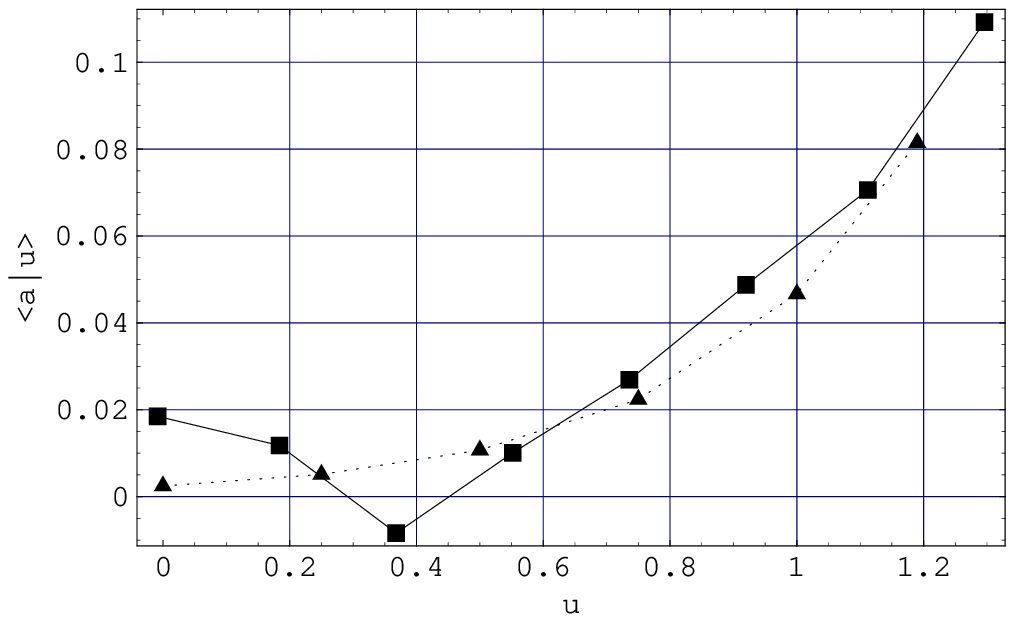}}
    \caption{\label{Fig2} The conditional mean acceleration
vs $u$. Triangles: $\langle a|u\rangle/\langle a^2|0\rangle^{1/2}$
for $\alpha=e^{|u|}$ and $\lambda=-0.005e^{3|u|}$. Boxes: the
experimental data on $\langle a|u\rangle/\langle
a^2\rangle^{1/2}$~\cite{Mordant0303003}.}
  \end{center}
  \end{minipage}
\end{figure}

The conditional PDF is given by (\ref{PLaval}) treated in the form
$P(a|\alpha(u),\lambda(u))$. The constant $C$ in (\ref{PLaval}) is
calculated for each value of $u$. For the normalized velocity
fluctuations values $u=0,\, 0.25,\, 0.50,\, 0.75,\, 1.00,\, 1.19$
the distributions $P(a|\alpha(u),\lambda(u))$ and mean
accelerations $\langle a|u\rangle/\langle a^2|0\rangle^{1/2}$ are
shown in Figs.~\ref{Fig1} and \ref{Fig2}. One observes a good
qualitative correspondence of the obtained {\em conditional PDFs}
(Fig.~\ref{Fig1}, shifted for clarity) with the experimental
curves (Fig.~6a in~\cite{Mordant0303003}). Both the variance and
skewness of $P(a|u)$ increase for bigger $|u|$. While the increase
of the variance (related to the increase of $\alpha(u)$) is
readily seen, the increase of the skewness (related to the
increase of $\lambda(u)$) results in a rather small change of the
shape of distribution despite the fact that $\lambda$ varies by
about two order of magnitude (from $|\lambda|$=0.005 to 0.18).
Such a change can be however readily seen in a plot of the
contribution to fourth order moment, $a^4P(a)$ (see Fig.~4
in~\cite{Aringazin6}). The obtained {\em conditional mean
acceleration} $\langle a|u\rangle/\langle a^2|0\rangle^{1/2}$
plotted in Fig.~\ref{Fig2} is also in a good qualitative agreement
with the experimental dependence $\langle a|u\rangle/\langle
a^2\rangle^{1/2}$ (Fig.~6b in~\cite{Mordant0303003}). The mean
acceleration is zero for a symmetrical distribution and for
homogeneous isotropic turbulence. One observes a rather small
relative increase of the mean acceleration for bigger $|u|$ that
eventually reflects a coupling of the acceleration to large scales
of the studied flow~\cite{Laval,Chevillard0310105}. We note that
the experimental $\langle a|u\rangle/\langle a^2\rangle^{1/2}$
exhibits small asymmetry with respect to $u\to -u$.

In summary, the presented 1D model is shown to capture main
features of the observed conditional acceleration statistics.




\end{document}